\documentclass[12pt,twoside,a4paper]{article}
\usepackage[dvips]{epsfig}
\usepackage{rotating}
\begin{document}

\begin{titlepage}
\vspace*{-3.0cm}
{\flushright LC-DET-2001-042 \\}

\vspace{1.0cm}

\begin{center}
{\huge \bf A Pixel Vertex Tracker\\ 

\vspace{0.15cm}

for the TESLA Detector}

\vspace{0.15cm}

\vspace{1.25cm}

{\large M.~Battaglia}\\
CERN, Geneva (Switzerland) and \\
Dept. of Physics, University of Helsinki (Finland)

\vspace{0.25cm}

{\large S.~Borghi and R.~Campagnolo}\\
Dip. di Fisica, Universita' degli Studi and INFN, Milano (Italy)

\vspace{0.25cm}

{\large M.~Caccia}\\
Dip. di Scienze, Universita' dell'Insubria and INFN, Como (Italy)

\vspace{0.25cm}

{\large K.~Domanski, P.~Grabiec, B.~Jaroszewicz,\\ 
J.~Marczewski, D.~Tomaszewski}\\
Institute of Electron Technology, Warszawa (Poland)

\vspace{0.25cm}

{\large W.~Kucewicz}\\
Dept. of Electronics, University of Mining and Metallurgy, Krakow (Poland)

\vspace{0.25cm}

{\large A.~Zalewska}\\
High Energy Physics Lab., Institute of Nuclear Physics, Krakow (Poland)

\vspace{0.25cm}

{\large K.~Tammi}\\
Helsinki Institute of Physics (Finland)

\end{center}

\vspace{0.75cm}

\begin{abstract}
In order to fully exploit the physics potential of a $e^+e^-$ linear collider,
such as TESLA, a Vertex Tracker providing high resolution track 
reconstruction is required. 
Hybrid Silicon pixel sensors are an attractive sensor technology option due 
to their read-out speed and radiation hardness, favoured in 
the high rate TESLA environment, but have been so far limited by the 
achievable single point space resolution. 
A novel layout of pixel detectors with interleaved cells to improve their 
spatial resolution is introduced and the results of the characterisation of 
a first set of test structures are discussed. In this note, a conceptual 
design of the TESLA Vertex Tracker, based on hybrid pixel sensors is presented.
\end{abstract}

\end{titlepage}

\section{Introduction}

The next generation of high energy $e^{+}e^{-}$ experiments,
following the LEP and SLC programs, will be at a linear collider.
The TESLA concept, based on super-conducting accelerating structures, will be
able to deliver luminosities in excess to 10$^{34}$~cm$^{-2}$~s$^{-1}$, 
at centre-of-mass energies ranging from 
the $Z^0$ pole up to about 1~TeV. Such a collider will naturally complement 
the physics reach of the Tevatron and LHC hadron colliders in the study of the
mechanism of electro-weak symmetry breaking and in the search for new physics 
beyond the Standard Model. Both precision measurements and particle searches 
set stringent requirements on the efficiency and purity of the flavour 
identification of hadronic jets since final states including short-lived $b$ 
and $c$-quarks and $\tau$ leptons are expected to be the main signatures. High
accuracy in the reconstruction of the charged particle trajectories close to 
their production point is required in order to reconstruct the topologies of 
the secondary vertices in the decay chain of short-lived heavy 
flavour particles.

If a light, elementary Higgs boson exists, as indicated by the LEP, SLD and 
LEP-2 data, it will be essential to carry out precision measurements of its 
couplings to the different fermion species as a proof of the mass generation 
mechanism and in order to identify its Standard Model or Supersymmetric 
nature~\cite{higgs}. 
This can be achieved by accurate determinations of its decay rate to 
$b\bar{b}$, $c\bar{c}$, $\tau^{+}\tau^{-}$, $W^{+}W^{-}$ and 
gluon pairs to detect possible deviations from the Standard Model 
predictions. Since the rates for the Higgs boson decay modes 
into lighter fermions $H^0 \rightarrow c \bar c$, $\tau^+ \tau^-$ or into 
gluon pairs are expected to be only about 10\% or less of that for the 
dominant $H^0 \rightarrow b \bar b$ process, the extraction and measurement 
of these signals require the suppression of the $b \bar b$ contribution by a 
factor of twenty or better, while preserving a good efficiency. 

The measurement of the Higgs-top Yukawa coupling as well as that 
of the top-quark mass will also require efficient $b$-tagging to reduce the 
combinatorial background in the reconstruction of the six and eight jet final
states. 
If Supersymmetry is realized in nature, the study of its rich Higgs sector will
also depend upon the efficient identification of $b$~jets and $\tau$ leptons 
to isolate the signals for the decays of the heavier $A^0$, $H^0$ and 
$H^{\pm}$ bosons from the severe combinatorial backgrounds in the complex 
multi-jet hadronic final states. 
Due to the expected large $b$~jet multiplicity, highly 
efficient tagging is required to preserve a sizeable statistics of the 
signal events. Finally, both $b$ and $c$-tagging will be important in the study
of the quark scalar partners, while $\tau$ identification may be instrumental 
in isolating SUSY signals in large $\tan\beta$ scenarios and Gauge 
Mediated Supersymmetry Breaking models. 

\begin{figure}[h!]
\begin{center}
\epsfig{file=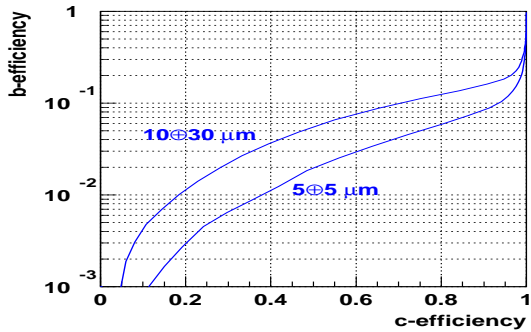,height=5.5cm,width=8.5cm,clip}
\caption{\sl The $b$-quark background efficiency as a function of the 
$c$-quark signal efficiency for a $c$-tag algorithm based on topological 
vertexing and impact parameters, developed on the basis of the experience 
from the SLC and LEP experiments (from~\cite{borisov}). 
The two curves show the expected performance for two different assumptions 
on the track impact parameter resolution.}
\label{fig:ctag}
\end{center}
\end{figure}

In order to correlate the track extrapolation accuracy to the performances 
of a jet flavour tagging algorithm at the linear collider,
Figure~1 shows the $c$ / $b$ quark separation capabilities  
under different assumptions for the track impact parameter resolution. 

While the developments of Vertex Trackers from the LEP and SLD experiments
to those at B-factories, the Tevatron and the LHC have addressed 
many crucial aspects, such as radiation hardness, read-out speed and 
the considerable size increase of the trackers, the impact parameter ($i.p.$) 
resolution, $\sigma_{ip}$, has not benefited of significant improvements. 
Limited in terms of multiple scattering at $B$ factories and by radiation 
damage in their closest approach to the beam at hadron colliders, the 
typical performances are comparable to those achieved at LEP: $\sigma_{ip}$ = 
25~$\mu$m$\oplus$70~$\mu$m/$p_t$~GeV/$c$. 
In order to fulfil its challenging requirements, the TESLA Vertex Tracker 
aims at significantly improving on the already outstanding SLD VXD3 
performances of 8~$\mu$m$\oplus$33~$\mu$m/$p_t$ GeV/$c$~\cite{vxd3}. 
An improvement to $\sigma_{ip}$ to  5~$\mu$m$\oplus$10~$\mu$m/$p_t$ GeV/$c$ 
corresponds to a factor $\simeq~2$ increase in charm jet tagging efficiency 
and a factor $\simeq 1.25$ increase in beauty jet tagging efficiency at 
constant mis-identification probability. A precise determination of the track 
perigee parameters close to their production point also assists in the track 
reconstruction and improves the momentum resolution for charged particles.   

The design of the TESLA Vertex Tracker and the choice of the sensor 
technology are driven by these requirements, to be achieved within the 
constraints set by the accelerator induced backgrounds at the interaction
region and by the characteristics of the physics events.
In the next section, the conceptual design proposed for the Vertex Tracker of 
the TESLA detector is presented while Section~3 reviews its expected 
performances.
In Section~4, the silicon pixel sensor design, developed to overcome
the hybrid pixel sensor limitations in terms of single point space resolution
is discussed in details. A road-map of dedicated R\&D activity, to 
validate the concepts presented, concludes. 

\section{Vertex Tracker Conceptual Design}

The proposed Vertex Tracker geometry consists of three layers of pixel 
detectors, providing at least three space points for charged particles with 
polar angles down to $|\cos \theta|$ = 0.995. In order to optimise the 
amount of material traversed by the particles and their angle of incidence on 
the sensitive plane, the outermost layer is split into a barrel section and a 
forward crown as shown in Figure~2. 
A {\sc Geant} model of this geometry has been implemented in 
the {\sc Brahms} simulation program~\cite{brahms} and used for its 
optimisation and the evaluation of the performances in terms of the impact 
parameter resolution as discussed in Section~4. The radial position of the 
innermost layer is set at 1.5~cm by the beam-pipe radius, limited to 1.4~cm due
to constraints on the design of the machine collimation system.
The outermost layer has been positioned at a radius of 10~cm, fitting the 
proposed general detector concept that includes two further silicon detector
layers located between the Vertex Tracker and the TPC main tracker. 
\begin{figure}[h!]
\begin{center}
\begin{tabular}{c c}
\epsfig{file=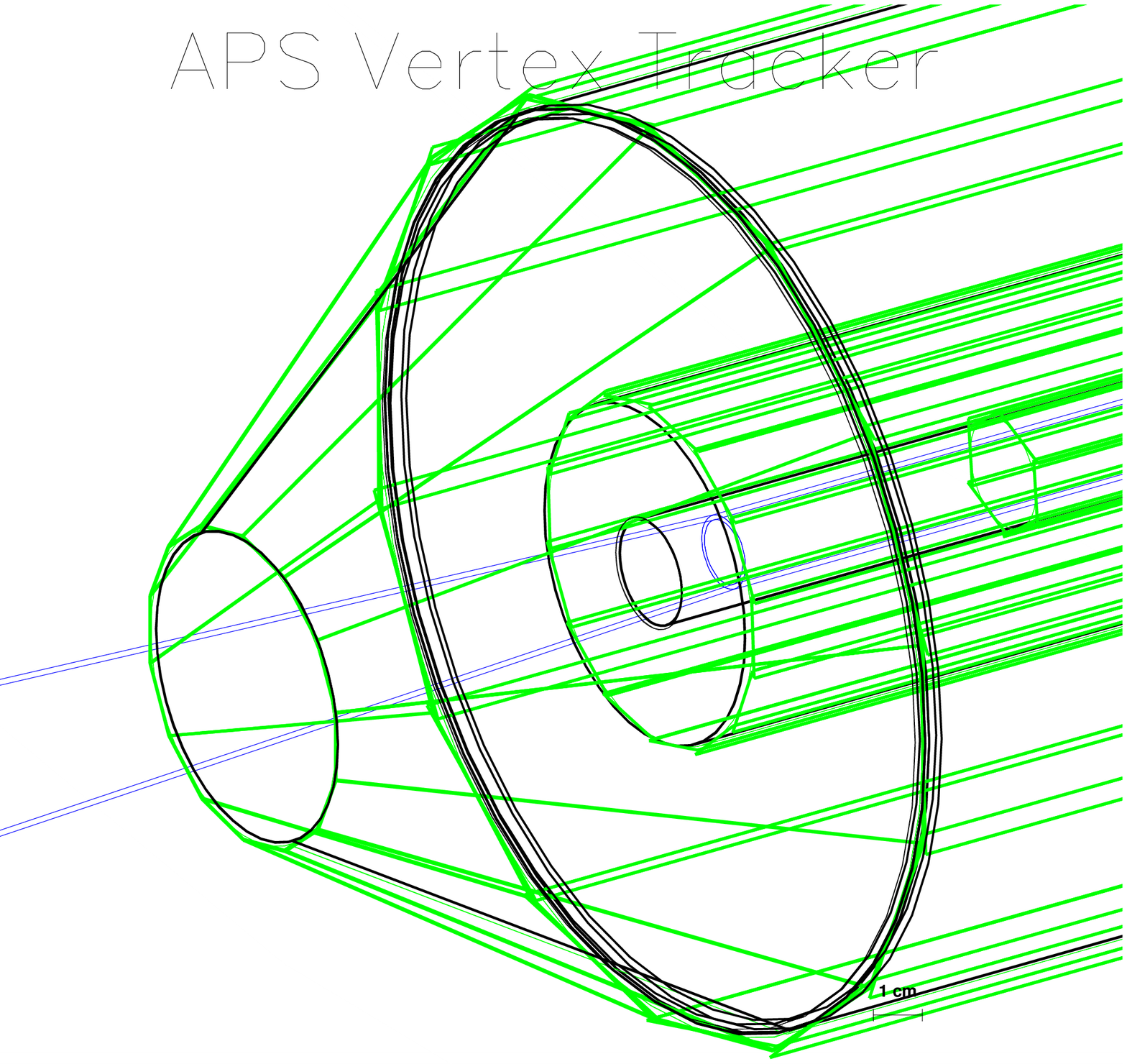,width=7.0cm,clip} &
\epsfig{file=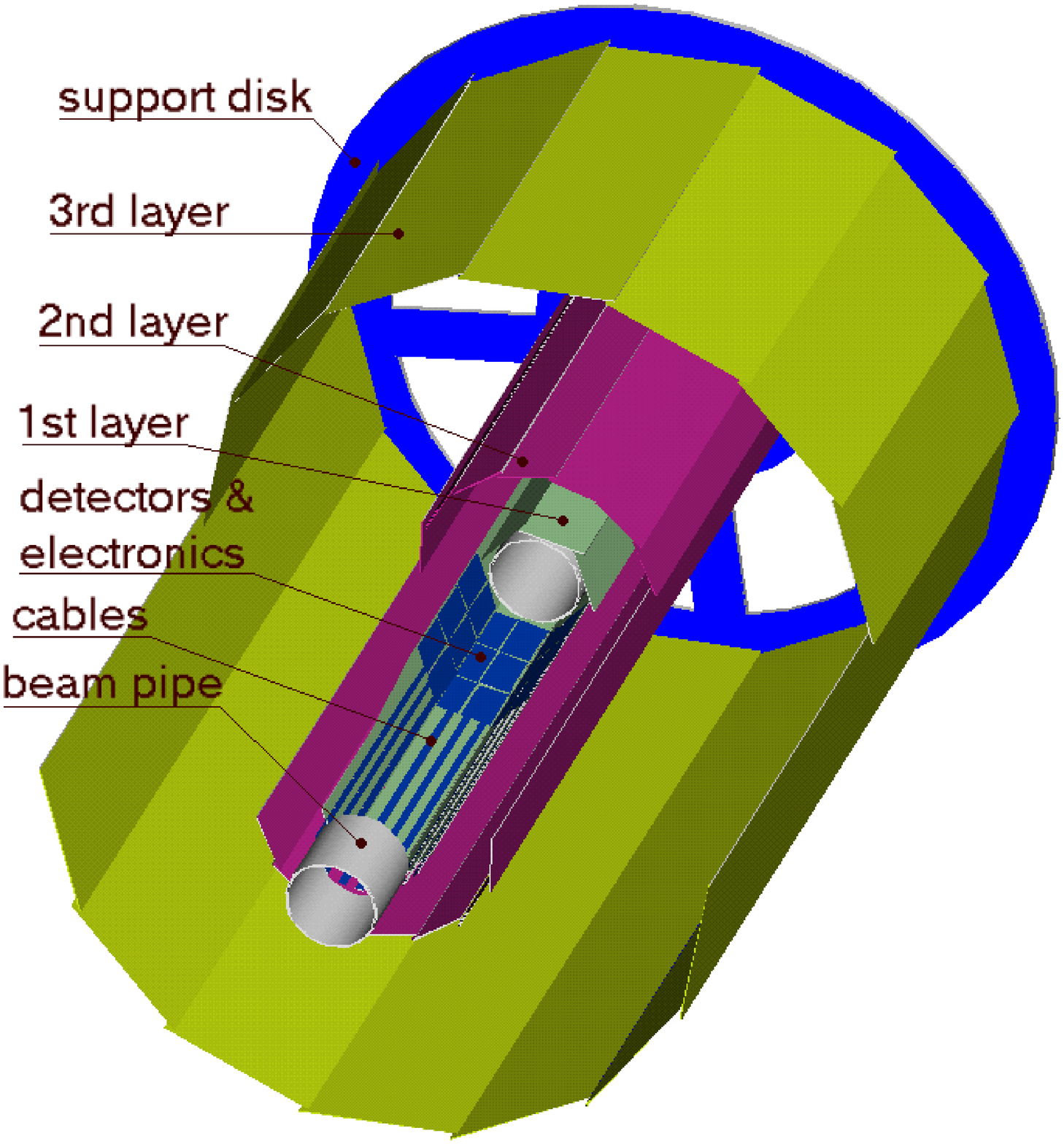,width=7.5cm,clip}\\
\end{tabular}
\end{center}
\caption{\sl A view of the proposed silicon Vertex
Tracker with three concentric layers as implemented in the {\sc Geant} (left) 
and {\sc Cad} (right) models. A forward crown of detector is 
attached at both ends on the carbon fibre support structure.}
\label{fig:geo}
\end{figure}
%The number and positions of the other layers have been optimised taking into 
%account also
%the occupancy due to the track density in the core of collimated hadronic jets
%and to the pair background. A layer efficiency of 97\% has been assumed. 
The number of additional intermediate layers is limited by the multiple 
scattering.
Detailed simulation studies, assuming a layer efficiency of 97\%, have shown 
that an optimal configuration consists of one intermediate layer located at 
3.5~cm, this layer assisting the extrapolation to the innermost layer 
affected by the higher occupancy and 
supplementing a space point close enough to the interaction point in cases of
failures on the first layer.
Overlaps of neighbouring detector modules provide an useful mean to verify
their relative alignment using particle tracks from dedicated 
calibration runs, taken at $Z$ centre-of-mass energy.

The limitation in the amount of material traversed by charged particles being a
major concern to minimise the multiple scattering effects, the pixel Vertex 
Tracker requires a light support structure able to cope with the load of the 
detector plaquettes and to ensure their stability to better than their 
intrinsic single point resolution. The proposed concept for the mechanical 
structure envisages the use of diamond-coated thin carbon fibre support 
shells, acting also as heat pipes to extract the heat dissipated by the 
read-out electronics, uniformly distributed over the whole active surface of 
the detector. A support plate at each end of the barrel section, made of two 
thin carbon fibre skins glued together by means of small spacers ensures the 
needed rigidity and provides a routing for the cables and a support for the 
forward crown (see Figure~3).  

\begin{figure}[h!]
\begin{center}
\begin{tabular}{c c}
\epsfig{file=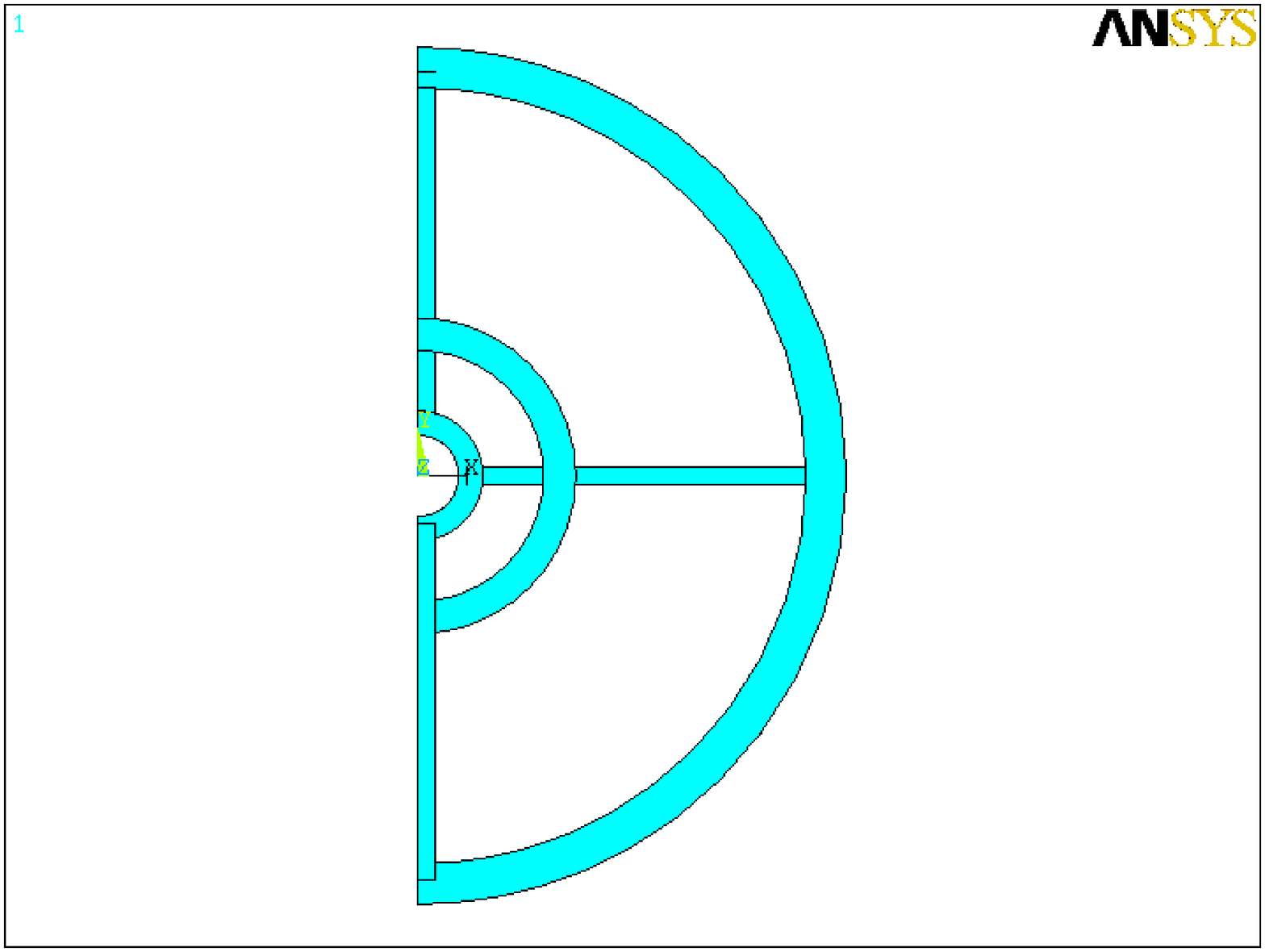,width=6.75cm,angle=270,clip} &
\epsfig{file=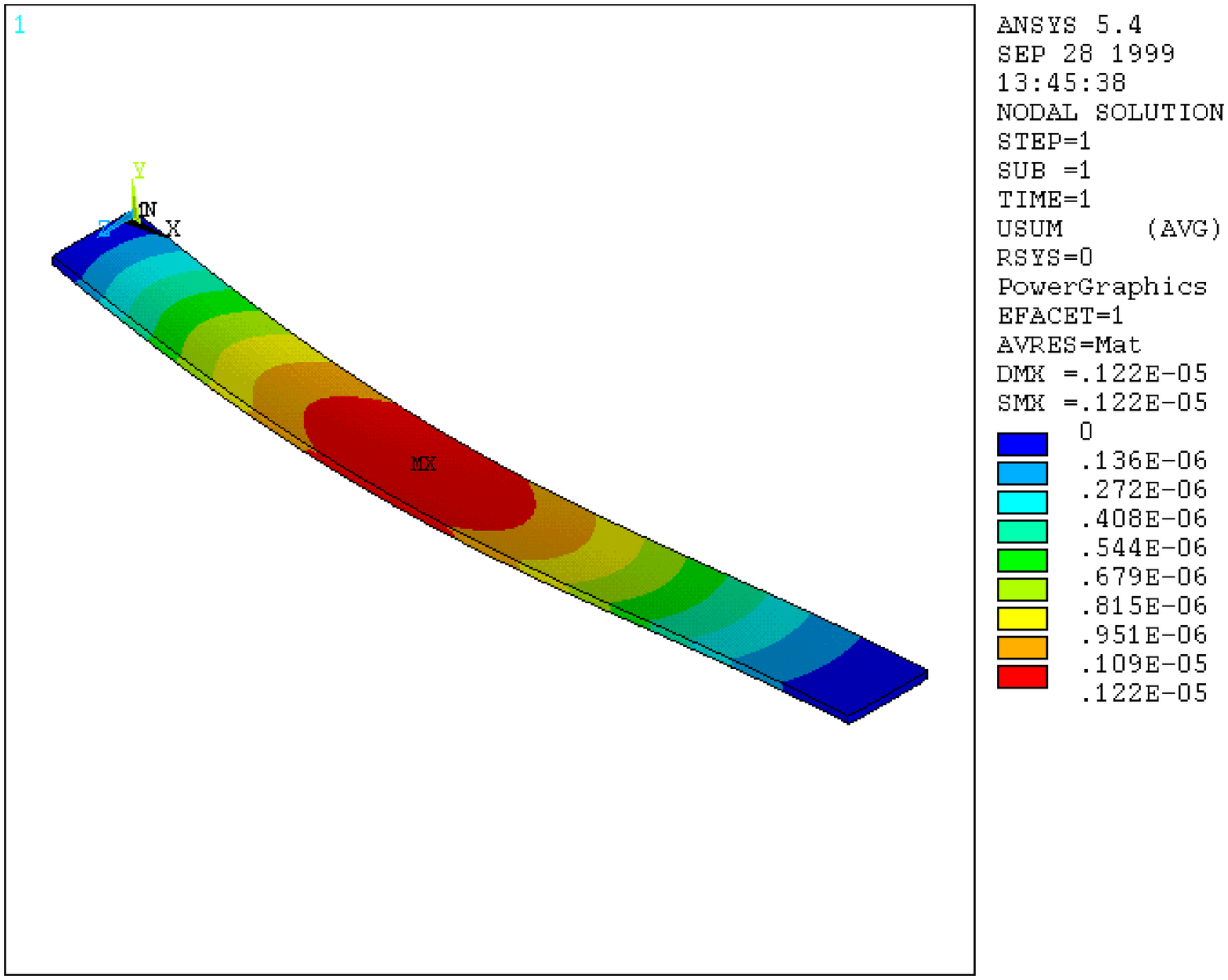,
width=6.75cm,angle=270,clip} \\
\end{tabular}
\end{center}
\caption{\sl Left: The carbon fibre end-plate support structure and 
Right: result of a finite element analysis of the deformations induced on a
detector module due to the temperature gradient. The module is fixed at 
both ends and the largest displacement corresponds to 10~$\mu$m for the 
innermost layer.}
\label{fig:ansys}   
\end{figure}

Assuming a power dissipation of 40~$\mu$W/channel, the total heat flux is 
790~W, corresponding to 2600~W/m$^2$, for a read-out pitch of 
100~$\mu$m. In presence of a power dissipation $Q$, the temperature 
gradient along the silicon sensor $\Delta T$ scales as  
$\Delta T = \frac{Q \ell^2}{2 k t}$, where $\ell$ is the distance, $t$ the 
sensor thickness and $k$ the thermal conductivity. Owing to its high thermal 
conductivity (1000-2000~W/m~K), small thermal expansion 
($\simeq 2 \times 10^{-6}$ $^{\circ}C$$^{-1}$), high electrical resistivity 
and large radiation length $X_0$, diamond is particularly well suited as a 
heat pipe material. 
The envisaged cooling scheme consists of a diamond coated heat pipe integrated
onto the support layer and heat drains regularly spaced. These drains,
consisting of pipes circulating liquid coolant, must be placed every 5~cm 
along the longitudinal coordinate, except for the innermost layer where they 
can be placed only at the detector ends to minimise the amount of material.
The low duty cycle of the accelerator, makes interesting to consider pulsed 
power operation to reduce power dissipation.

Signals can be routed along the beam pipe and the end-cap disks to the 
repeater electronics installed between the vertex detector and the forward
masks which shield the detector from background.
The material budget has been estimated by assuming 200~$\mu$m thick detectors 
and back-thinning of the read-out chip to 50~$\mu$m, corresponding to 
0.3~\%~X$_{\rm0}$, and the support structure. It corresponds to 
1.6~\%~X$_{\rm0}$ for the full tracker (see Figure~4 and 
Table~1). 
\begin{figure}[h!]
\begin{center}
\epsfig{file=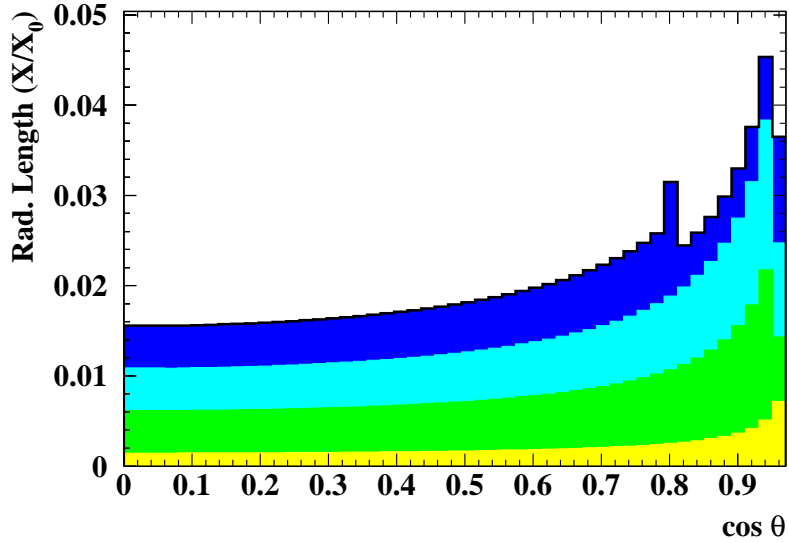,width=11.0cm,clip}
\end{center}
\caption{\sl The material budget, in units of radiation length, for the 
proposed pixel Vertex Tracker as a function of the polar angle. The 
contribution from the Be beam-pipe and those of the three layers are shown 
from bottom to top respectively. The adoption of a forward crown of detectors
to complement the barrel section of the tracker allows to minimize the rise
of the material traversed by particles produced at the interaction region at
low angles.}
\label{fig:mat}
\end{figure}

\begin{table}[h!]
\begin{center}
\begin{tabular}{|l|c|c|c|c|c|}
\hline
\mbox{ } & $R$  & Material & $X$  & $X_0$  & $X/X_0$ \\
\mbox{ } & (cm) & \mbox{ } & (cm) & (cm$^{-1}$) & (\%) \\ 
\hline \hline
Beam-pipe        & ~1.4     & Be       & 0.0500   & 35       & 0.15  \\
Sensor+VLSI+Kapton  & ~1.5     & Si       & 0.0250   & 9.4      & 0.37  \\
Support   & ~1.2  & CF      & 0.0240   & 25                  & 0.10  \\
Sensor+VLSI+Kapton  & ~3.5     & Si       & 0.0250   & 9.4      & 0.37  \\
Support   & ~3.5  & CF      & 0.0240   & 25                  & 0.10  \\
Sensor+VLSI+Kapton  & 10.0     & Si       & 0.0250   & 9.4      & 0.37  \\
Support          & 10.0    & CF       & 0.0240   & 25       & 0.10  \\ \hline
Total     & \mbox{ } & \mbox{ } & \mbox{ } & \mbox{ }       & 1.60  \\
\hline
\end{tabular}
\end{center}
\caption{\sl The material budget for the pixel Vertex Tracker with the 
detailed contributions from its different components.}
\label{tab:mat}
\end{table}

The expected data size from the Vertex Tracker has been computed assuming 
600 hits per bunch crossing from background sources and 50 charged particles. 
Taking 16~bits per pixel, this corresponds to 150 kbyte of data from the 
Vertex Tracker per $e^+e^-$ event.

\section{Simulation Studies and Vertex Tracker Performances}

The performances of the pixel Vertex Tracker have been evaluated for 
$e^+e^- \rightarrow HZ \rightarrow q \bar q \ell^+\ell^-$ and 
$e^+e^- \rightarrow t \bar t \rightarrow W^+ b W^- \bar b$ events at 
$\sqrt{s}$ = 350 and 500~GeV using full {\sc Geant} simulation. Tracks 
reconstructed in the TPC {\sc Tesla} main tracker have been extrapolated to the
detector planes of the pixel Vertex Tracker and a dedicated pattern 
recognition program has been used to define the association of the recorded 
hits to the extrapolated tracks. While no attempt has been made to define 
local track element using only the three layers of the Vertex Tracker, these 
can be efficiently defined by adding the hits on the two layers of the Silicon
Intermediate Tracker, foreseen at radii of 16 and 30~cm. The impact parameter
resolution has been studied as a function of the Vertex Tracker geometry, 
namely the radial position of the second layer, and of the assumed single 
point resolution $\sigma_{point}$. Results are summarised in 
Tables~2 and~3.
\begin{table}[h!]
\begin{center}
\begin{tabular}{|c|c|}
\hline
$\sigma_{point}$ ($\mu$m) & $\sigma^{asymptotic}_{IP}$ ($\mu$m) \\
                          & ($R-\phi$) \\ \hline \hline
~4. & 3.7 \\
~7. & 5.3 \\
10. & 7.5 \\
13. & 9.5 \\  \hline
\end{tabular} 
\end{center}
\caption{\sl Observed variation of the asymptotic impact parameter 
resolution for different values of the assumed detector single point 
resolution.}
\label{tab:ip1} 
\end{table}
\begin{table}[h!]
\begin{center}
\begin{tabular}{|c|c|c|}
\hline
$R_2$ & Asymptotic $\sigma_{IP}$   & Fraction of \\
(cm)  & ($R-\phi$) ($\mu m$)     & Tracks in Tails\\
\hline \hline
7.5   & 6.0   &  12 \% \\
5.5   & 5.6   &  ~7 \%   \\
3.5   & 5.2   &  ~4 \%   \\ \hline
\end{tabular}
\end{center}
\caption{\sl Observed variation of the asymptotic impact parameter 
resolution and the fraction of tracks in the non-Gaussian tails of the 
resolution function for different values of the assumed radial position $R_2$ 
of the second Vertex Tracker layer.}
\label{tab:ip2} 
\end{table}
The efficiency for associating at least two hits to an extrapolated charged 
particle with $p >$~1~GeV/$c$ has been found to be $\simeq 95\%$. The track
fit has been performed using a Kalman-filter~\cite{trkfit} and the impact
parameter resolution has been obtained from the estimated covariance matrix.
The results obtained assuming $\sigma_{point}$ = 7~$\mu$m, are shown in 
Figure~5.

\begin{figure}[h!]
\begin{center}
\epsfig{file=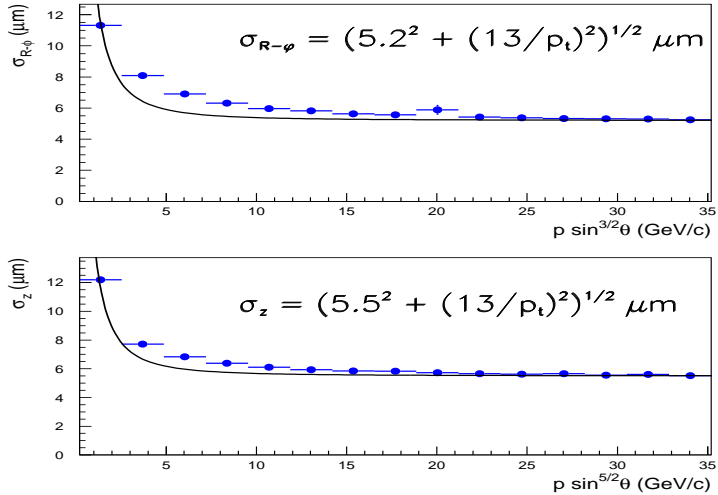,width=10.5cm,height=7.5cm,clip}
\end{center}
\caption{\sl Impact parameter resolution in the $R-\phi$ (upper plot) and 
$R-z$ (lower plot) projections for particles as a function of their 
momentum and polar angle.}
\label{fig:ip}
\end{figure}

\section{Hybrid Pixel Sensors}

The main requirements for the Vertex Tracker sensors may be summarised as
follows:
\begin{itemize}
\item A single point resolution $\le$ 7 $\mu$m and a thickness of 
      $\le 200$~$\mu$m to attain the required impact parameter resolution.
\item Time stamping capabilities to identify particles originating from a 
      single bunch crossing, occurring at TESLA every 337 (189)~ns at
      $\sqrt{s}$ = 500 (800)~GeV to reduce the pair
      background and the number of $\gamma \gamma \rightarrow 
      {\mathrm{hadrons}}$ events overlapped to the products of a $e^+e^-$ 
      collision.
\item A sensitive cell area below ${\rm 150 \times 150 \mu m^2}$ to keep the 
      occupancy from backgrounds and dense hadronic jets below 1\%.
\end{itemize}
At present, two silicon sensor technologies 
have the potential to satisfy these specifications:
monolithic (Charged Coupled Devices (CCD) and CMOS imagers)
and hybrid pixels sensors. 
Hybrid pixel sensors have the advantage of allowing fast time stamping 
and sparse data scan read-out, thereby reducing the occupancies due to
backgrounds, and of being tolerant to neutron fluxes well beyond 
those expected at the linear collider. Both these characteristics
have been demonstrated for their application in the LHC experiments. 
However, improvements in the pixel sensor spatial resolution and a reduction 
of the sensor total thickness are needed. These represent areas of R\&D that 
are specific to this linear collider application.
CCD require a complementary development. Their readout speed and radiation 
tolerance need to be improved, their resolution and thickness being already 
close to requirements. CMOS imagers have recently appeared and could overcome 
the CCD limitations while offering similar advantages.

A single point precision at the 5~$\mu$m level can be obtained by 
sampling the diffusion of the carriers generated along the particle path and 
adopting an analog read-out  to interpolate the signals of 
neighbouring cells. 
Since the charge diffusion r.m.s. is $\approx$~8~$\mu$m in 300~$\mu$m thick 
silicon, an efficient sampling requires a pixel pitch well below 
50~$\mu$m.
At present, the most advanced pixel read-out electronics have minimum 
cell dimension of ${\rm 50 \times 300 \mu m^2}$, limiting an efficient signal 
interpolation. Even if a sizeable reduction in the cell dimension 
may be envisaged by the deep submicron trend in the VLSI development, 
a sensor design overcoming such basic limitation is definitely worth 
being explored.

\subsection{Detector design and first prototype production}

The pixel detector design discussed here exploits  a layout inherited 
from the microstrip detectors \cite{hyams} where 
it is assumed to have a read-out pitch n times larger than the 
pixel pitch. The proposed  sensor layout is shown in Figure~6 
for n=4. In such a configuration, the charge carriers created underneath an 
interleaved pixel  will induce a signal on the output nodes, 
capacitively coupled to the interleaved pixel. 
\begin{figure}[h!]
\begin{center}
\epsfig{file= 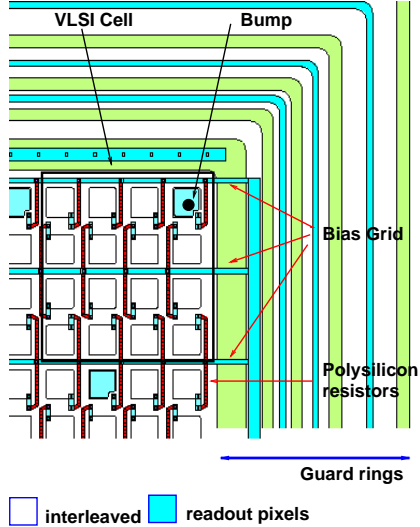,height=7.0cm,clip}
\caption{\sl Layout of the proposed detector, corresponding to a 
${\rm 50~\mu m}$ pitch in both dimensions; the other structures differ by the 
pitch only}
\label{fig:corner}
\end{center}
\end{figure}
In a simplified model where the detector is reduced to a 
capacitive network, the ratio of the signal amplitudes on the 
output nodes at the left hand side  and right hand  side of  
the interleaved pixel (in both dimensions) should have a linear 
dependence on the particle position. 
The ratio between the inter-pixel capacitance and the pixel 
capacitance to backplane plays a crucial role in the detector design, 
as it defines the signal amplitude reduction at the output nodes and therefore
the maximum sustainable number of interleaved pixels. 
Recently published results on ${\rm 200~\mu m}$ readout pitch microstrip 
detectors have demonstrated a ${\rm 10~\mu m }$ resolution for a layout with 
three interleaved strips (${\rm 50~\mu m}$ strip pitch) and 
${\rm S/N \approx 80}$~\cite{krammer}. Similar results may be expected in a 
pixel detector, taking into account that a lower noise is achievable due to 
the intrisically smaller load capacitance and that the charge sharing on 
four output nodes makes possible to reconstruct the particle position in two 
dimensions. Further improvements are possible by sampling the charge carrier 
diffusion at a smaller pitch.   

Prototypes of detectors with interleaved pixels have been designed and 
manufactured. The layout of one of the structures is shown in 
Figure~6. A series of guard rings defines the
detector sensitive area. A bias grid allows the polarization of the interleaved
pixels too; each ${\rm p^{+}}$ implant is connected to the metal bias line
 by polysilicon resistors in the  ${\rm 1-3~M\Omega}$ range.
A metal layer is deposited on top of the pixels to be connected to the
VLSI cell. The backplane has a meshed metal layer to allow the use of an IR
diode for charge collection studies. In a 4", 350~$\mu$m thick, wafer 
36 structures were fit, for 17 different layouts; 
a VLSI cell of ${\rm 200 \times 200~\mu
m^{2}}$ or ${\rm 300 \times 300~\mu m^{2}}$ was assumed and detectors with a 
number of interleaved pixels ranging between 0 and 3 and different areas 
were designed. 

Ten high resistivity wafers ${\rm (5-8~k\Omega cm)}$ were processed together 
with an equal number of low resistivity wafers for process control~\cite{iet}.
The technological sequence may be outlined as follows \cite{az}:
\begin{itemize}
\item wafer oxidation by HCl gettering, to form a 700 nm thick Silicon dioxide
layer;
\item Phosphorus glass deposition from a ${\rm POCl_3}$ source followed by 
drive-in in oxidizing environment. The P diffusion was meant to form an 
${\rm n^+}$ layer on the backplane, to guarantee a good ohmic contact with the
metal grid;
\item pixel and guard ring window opening in the photomask, applying a wet 
etching technique;
\item Boron diffusion, followed by a drive-in process. The resulting junctions
were ${\rm 0.7\mu m}$ deep;
\item high quality thin oxide (200 nm) growth at ${\rm 1000~^{o}C}$ to define 
the integrated coupling capacitor on every pixel. This step was carefully 
monitored to achieve an extremely low density of mobile ion charge;
\item polysilicon layer deposition (500 nm thick) and implantation, for the 
bias resistors. A special high temperature annealing was optimised to obtain 
the required sheet resistence ${\rm 250 k \Omega}$;
\item resistor pattern definition by photolitography;
\item planarisation by BPSG (Boron-Phosphorus-Silicon Glass) chemical vapour 
deposition, followed by a high temperature annealing;
\item resistor coating by photolitography, applying plasma and wet etching 
technique;
\item definition of the contact window between the resistors and the 
${\rm p^+}$ implants. To decrease the contact resistence, Boron implantation 
was used;
\item metal deposition by sputtering a ${\rm 1.2 \mu m}$ thick Al-Si-Cu layer,
patterned by photolitography and plasma etching;
\item backside metallisation grid by Al evaporation and plasma etching
\item final wafer sintering in Hydrogen at ${\rm 450~^{o}C}$ for 20'.
\end{itemize}

\begin{figure}[h!]
\begin{center}
\epsfig{file=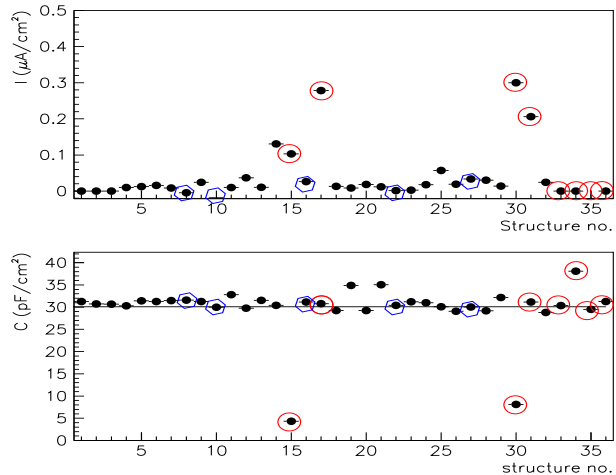,width=9.0cm,height=6.5cm,clip}
\caption{\sl Leakage current and capacitance at full depletion voltage for all 
of the 36 structures in one of the good wafers; circles define the detectors 
rejected because of breakdowns below ${\rm 100~V}$; hexagons identify rejected
detectors because of interrupted metal lines. The line corresponds to the 
expected capacitance per unit area.}
\label{fig:waf05}
\end{center}
\end{figure} 

\begin{figure}
\begin{center}
\epsfig{file=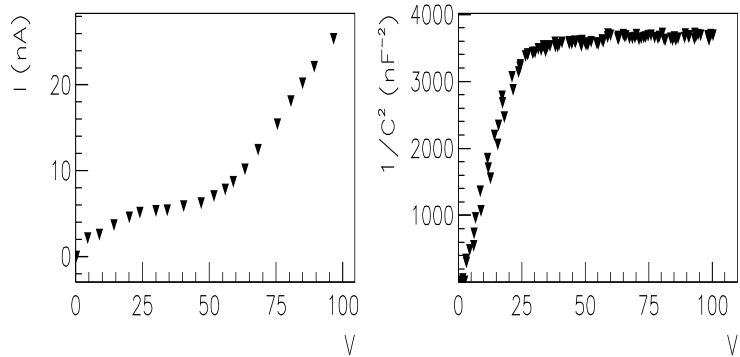,width=11cm,height=6.0cm,clip}
\caption{\sl ${\rm I~vs.~V}$ and ${\rm 1/C^{2}~vs.~V}$ curves for a typical 
good detector}
\label{fig:ivcv}
\end{center}
\end{figure}

Standard cleaning was performed before each high temperature step.
After the processing, two wafers were retained by the factory for a 
destructive analysis and two others were stored for later use.
All of the structures on the available
wafers were visually inspected, tested up to ${\rm 250~V}$ and 
characteristics I-V and C-V curves produced. In Figure~7 the 
value of the currents and detector capacitances at depletion voltage are 
shown for all of structures in one of the wafers. In Figure~\ref{fig:ivcv}
the typical ${\rm I~vs.~V}$ and ${\rm 1/C^{2}~vs.~V}$ curves of a good 
structure are shown. 
While the ${\rm C~vs.~V}$ curves behave as expected, the current has a 
peculiar trend. After a plateau is reached at full depletion, the current 
increases significantly at $V = 50-70$~V. For most of the
structures this is a mild increase, but most of the rejected detectors are 
characterized by a steep slope, eventually resulting in a breakdown for values
of the reverse bias voltage below 100~V. 
Independent measurements of the guard ring and bias grid current 
have shown the latter to be responsible for this effect, possibly due to sharp
edges where the electric field achieves high values. 
The average yield over the tested wafers is about 50\%; most of the rejected 
structures feature either an early breakdown or an extremely high leakage 
current, independent of the applied voltage, correlated to surface defects 
possibly associated to the plasma etching of the Al pattern. An improvement in 
the technology sequence, requiring a better planarisation and avoiding plasma 
etching has been agreed with the manufacturer. Good structures are 
characterised by a leakage current at the ${\rm 10 nA/cm^2}$ level and 
backplane capacitances in agreement to the expectations.

\subsection{Interpixel and backplane capacitance simulation\\ and measurements}

In order to study the charge collection properties of the proposed detector, a 
detailed characterization of its electrostatic properties is necessary.
In a simplified model, the detector may be reduced to a 
capacitive network. Each pixel is a node characterised by the backplane 
capacitance ($C_{bp}$) and the inter-pixel capacitances ($C_{ip}$), 
dominated by the couplings to the nearest and diagonal neighbours.
The $C_{ip}/C_{bp}$ ratio is crucial in the detector design, as it defines 
the signal amplitude reduction (an effective charge loss) at the 
output nodes. In order to be able to evaluate the expected behaviour of the
prototypes, the capacitances were measured and calculated. Moreover, the 
obtained values were used as input for a network analysis providing an 
estimate of the charge loss. 
Since single pixel capacitances are expected to be 
at the 10~$fF$ level, the measurement conditions were optimised as follows:

\begin{itemize}
\item The capacitance was measured using a Hewlett-Packard 4280 CV meter,
 operating at 1~MHz. A cable correction procedure accounting for finite 
 admittances to ground was used. Moreover, an open circuit correction
  was applied, measuring the capacitance with raised probe tips.
\item All pixels along a single metal bias line were set in parallel, 
isolating the line from the bias grid (see Figure~6). 
The number of pixels along a line ranges between 64 and 254, 
depending on the pixel pitch, thus increasing the measured 
capacitance values to 0.5-2~$pF$.
\item For every structure, three bias lines were isolated and the capacitance 
for each of them, for three doublets in parallel and for one triplet was 
measured. The offset in the linear regression for the capacitance versus the 
number of lines measured the left-over parasitic contributions after the cable 
correction.
\item For each structure, independent measurements of  the total interpixel 
capacitance and of the sum of the interpixel and backplane capacitance were 
done.
\end{itemize}

The numerical estimate is 
essential to break down the measurements in the single inter-pixel 
contributions, used to specify the capacitive network. The detector was simply 
modelled as an electrode matrix facing a metal plane, respecting the pitch,
implant width and dector thickness of the different prototypes. The 
capacitances
were obtained solving the Laplace equation with suitable boundary conditions.
In fact, setting 1~V on the central pixel 
and grounding the neighbouring pixels and the backplane, the stored charge on 
each pixel numerically equals its 
capacitance with respect to the central pixel, having a backplane capacitance
numerically equals to  the stored charge on the backplane. The Laplace 
equation was solved in a ${\rm 5 \times 5 }$ pixel
matrix, surrounded by a guard ring, with a finite element analysis performed
using the OPERA-3D package \cite{TOSCA}. The mesh needed by the equation solver
was optimised accounting for the physics 
aspects and the characteristics of the finite element procedure:
\begin{itemize}
\item because of the boundary conditions, in our model the potential 
      should always be positive
\item  the charge neutrality should not be violated, i.e. the sum of 
      the charge on all pixels and on the backplane should be zero.
\item a finer mesh had to be foreseen in the regions with higher 
$grad~{\bf E}$.
\end{itemize}
The final optimisation was based on the minimisation of the charge neutrality
violation; moreover, this was assumed as a measurement of the uncertainty on 
the simulation results and defines the errors in the summary table. By 
changing the the mesh characteristics in the high $grad~{\bf E}$ region, the 
net effect was estimated to be at the 10\% level. 
The results of measurements and simulation are reported in 
Table~4.

\begin{table}[h!]
\begin{center}
\begin{tabular}{|l|c|c|c|c|}
\hline
   & chip~1 & chip~2 & chip~3 & chip~4 \\
\hline \hline
Implant width [$\mu$m]     & 100 &  60 &   50 &   34 \\
Implant pitch [$\mu$m]     & 150 & 100 &   75 &   50 \\ 
Readout pitch [$\mu$m]     & 300 & 200 &  300 &  200 \\
No. of pixels in parallel  &  64 & 128 &  126 & 254 \\ \hline
Measured   $C_{ip}$  $[fF]$ & 599$\pm$5 & 1038$\pm$11 & 958$\pm$5 & 
2098$\pm$30\\
Calculated $C_{ip}$  $[fF]$ & 630$\pm$60 & 880$\pm$67 & 690$\pm$70 & 
980$\pm$40\\
Measured   $C_{bp}$ (I) $[fF]$ & 376$\pm$40 & 390$\pm$50 & 160$\pm$45 & 
314$\pm$35\\
Measured   $C_{bp}$ (II)  $[fF]$ & 447$\pm$5 & 368$\pm$5 & 218$\pm$5 & 
185$\pm$5\\
Calculated $C_{bp}$  $[fF]$ & 470$\pm$70 & 410$\pm$90 & 230$\pm$35 & 
211$\pm$60\\ \hline
\end{tabular} 
\end{center}
\caption{\sl Interpixel ($C_{ip}$) and backplane ($C_{bp}$) capacitance 
values for different detector structures. The reported values refer to the 
specified number of pixels in parallel, corresponding to a double column of 
pixels, to the left and right hand side of a single bias line}
\label{tab:capi}
\end{table}

The consistency of the measurements is checked comparing the value obtained 
subtracting the measured interpixel capacitance from the measurement of 
${\rm C_{ip} + C_{bp}}$ (measurement I in Table~4)
and the backplane capacitance from the asymptotic value in the CV curves, 
normalized by the number of pixels in the matrix (measurement II). 
The measurement cross check is satisfactory and the comparison with the 
calculated values is fair for all of the structures but chip~4,
where difficulties both in the measurement and in the simulation 
were expected because of the small pitch. In particular, the maximum
number of elements was preventing an optimal interpixel mesh and 
the extension to a larger pixel matrix, crucial for a proper interpixel 
capacitance evaluation of pixels with 50~$\mu$m pitch in a
350~$\mu$m thick detector. 
The calculated single pixel main capacitances are summarised in 
Table~5, together with the maximum charge loss resulting  by a 
network analysis with a dedicated software based on the node potential 
method~\cite{jacek}. The role of the interpixel capacitance $C_{ip}$ may be 
stressed referring to chip~4, where the maximum charge loss is reduced to 
51\% if the measured $C_{ip}$ values are taken. 

\begin{table}[h!]
\begin{center}
\begin{tabular}{|l|c|c|c|c|}
\hline
   & chip~1 & chip~2 & chip~3 & chip~4 \\
\hline \hline
Total $C_{ip}$ [$fF$] & 26.2$\pm$4.0 & 13.4$\pm$0.9 & 13.8$\pm$2.0 & 
9.6$\pm$1.0\\
$C_{nn}$  [$fF$] & 4.4$\pm$0.7 & 2.0$\pm$0.1 & 2.1$\pm$0.3 & 
1.5$\pm$0.2\\
$C_{bp}$ [$fF$] & 7.3$\pm$1.1 & 3.2$\pm$0.7 & 1.9$\pm$0.7 & 
0.8$\pm$0.2\\ \hline
Max. charge loss & 35\%  & 32\% & 77\%  & 67\% \\ 
\hline
\end{tabular} 
\end{center}
\caption{\sl Single interpixel ($C_{ip}$), to nearest neighbour pixel 
($C_{nn}$) and backplane ($C_{bp}$) capacitance values for different detector 
structures and estimated maximum charge loss obtained by modelling the 
detector as a capacitive network.}
\label{tab:cap2}
\end{table}

\subsection{Charge collection studies}
The achievable resolution is proportional to the pixel pitch and the Noise 
over Signal ratio (N/S). If the implant pitch is comparable to the diffusion 
width, the collected charge at the output nodes will have both the 
contributions by the direct diffusion and the linear share by the capacitive 
coupling. In such a case, an improvement beyond the binary value given by the 
$implant~pitch/\sqrt{12}$ may be expected. The N/S in pixel detectors can 
easily exceeds 1/100, due to the small detector capacitance. Because of this, 
an effective signal interpolation may be expected even with  $\simeq 50\%$ 
signal reduction. 

The charge collection properties and the achievable resolution have been 
directly studied by shining an infrared diode spot on the backplane of a 
structure with 60~$\mu$m implant width, 100~$\mu$m implant pitch and 
200~$\mu$m read-out pitch. At the diode wavelength of $\lambda$ = 880~nm, the 
penetration depth in the silicon substrate corresponds to 10~$\mu$m. The IR 
light has been focused to a spot size of $\simeq$~80~$\mu$m and its position 
in the detector plane controlled by a 2-D stage allowing to scan the pixel 
array with micro-metric accuracy. 
A sketch of the tested structure and the scan direction is displayed in 
Figure~9.
A matrix of 4 $\times$ 7 read-out pixels has been wire-bonded to a VA-1 
chip~\cite{ideas}. For each spot position, 1000 events have been recorded. 
The common mode, pedestal and noise calculation has been initialised for the 
first 300 events. In the subsequent events light was injected every 10 events,
allowing for continuous pedestal tracking. 
Because of the limited data volume, no on-line suppression has been applied 
and the data reduction and cluster search has been performed off-line. Results
have been averaged over the 70 recorded light pulses, with a peak pulse height
corresponding to N/S $\simeq$ 1/100 and a maximum charge loss of $\simeq$ 
40\%, in agreement with the network analysis. 
The charge sharing may be characterised by the $\eta$ function, 
defined as \mbox{$\eta = \frac{PH_i}{PH_{cluster}}$},
where $PH_i$ is the pulse height on 
the reference pixel $i$, normalised to the cluster pulse height.
\begin{figure}[ht!]
\begin{center}
\epsfig{file=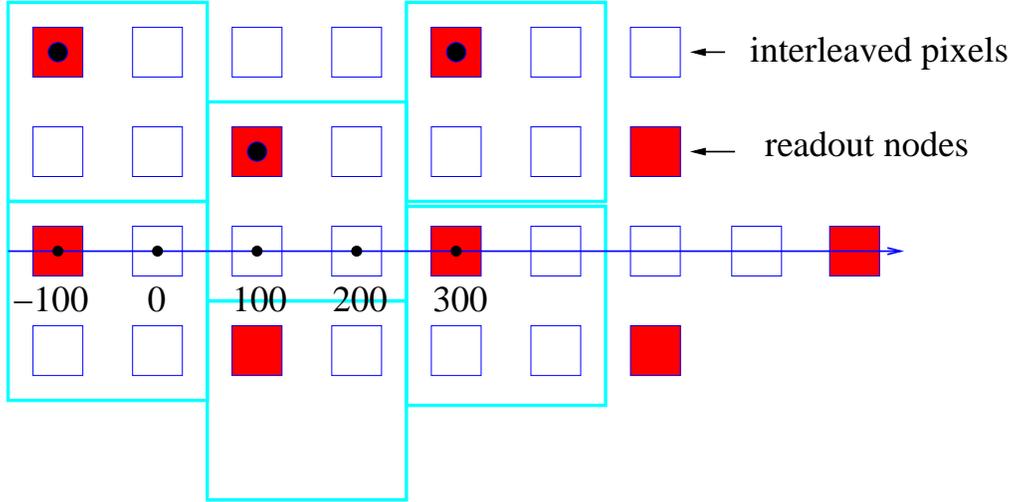,height=6.75cm,clip}
\caption{\sl A sketch of the tested structure, with 100~$\mu$m implant pitch 
and 200~$\mu$m read-out pitch. The horizontal line identifies the scan 
direction and defines the coordinate system. The frames define the footprint 
of an hypothetical VLSI cell.}
\end{center}
\label{fig:fnal}
\end{figure}
\begin{figure}[hb!]
\vspace*{-0.5cm}
\begin{center}
\epsfig{file=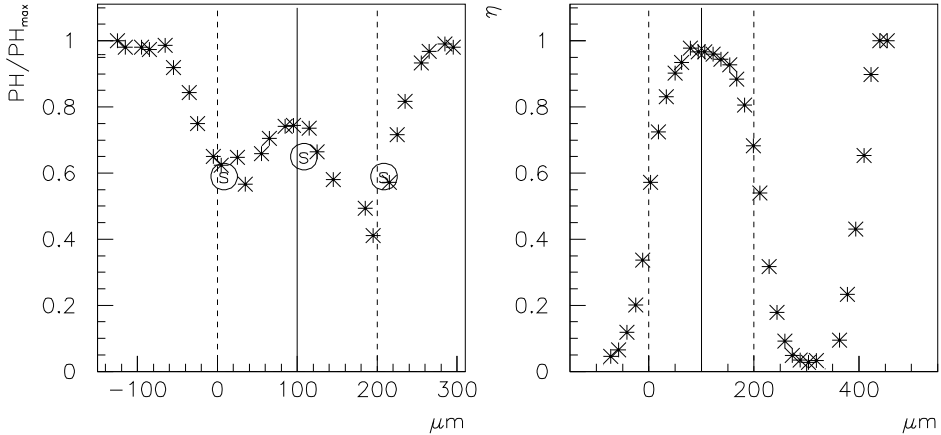,bbllx=14,bblly=250,bburx=532,bbury=524,height=6.75cm,
clip}
\caption{\sl The pulse height (left) and the charge sharing among neighbouring 
read-out pixels (right), measured by $\eta = \frac{PH_i}{PH_{tot}}$, during 
the detector scan. The vertical lines identify the centre of the pixels; the 
full line indicates the reference pixel used for the $\eta$ calculation. 
In the left plot, the encircled "s" marks represent the predictions from the 
simulation.}
\end{center}
\label{fig:eta}
\end{figure}

The $\eta$ function by construction ranges in the [0;1] interval and it has a 
period equals to the readout pitch. The measured pulse height and ${\rm \eta}$ 
distributions are shown in Figure~10, where 
the reference pixel for the first period has a centre at x = 100~$\mu m$.
For the structure under test, the ratio between the spot size and the pitch 
$\simeq$ 0.8 and the experimental $\eta$ curve 
can be understood as a superposition of the effects due to the 
diffusion of the charge carriers created by the IR spot on the 
neighbouring junctions and by the capacitive charge sharing. The $\eta$ 
parametrisation allows a coordinate reconstruction on an event by event basis 
and the measurement of the resolution, obtained comparing the laser spot 
position by the micro-metric stage to the reconstructed values. The results are
shown in Figure~11.

\begin{figure}[ht!]
\begin{center}
\epsfig{file=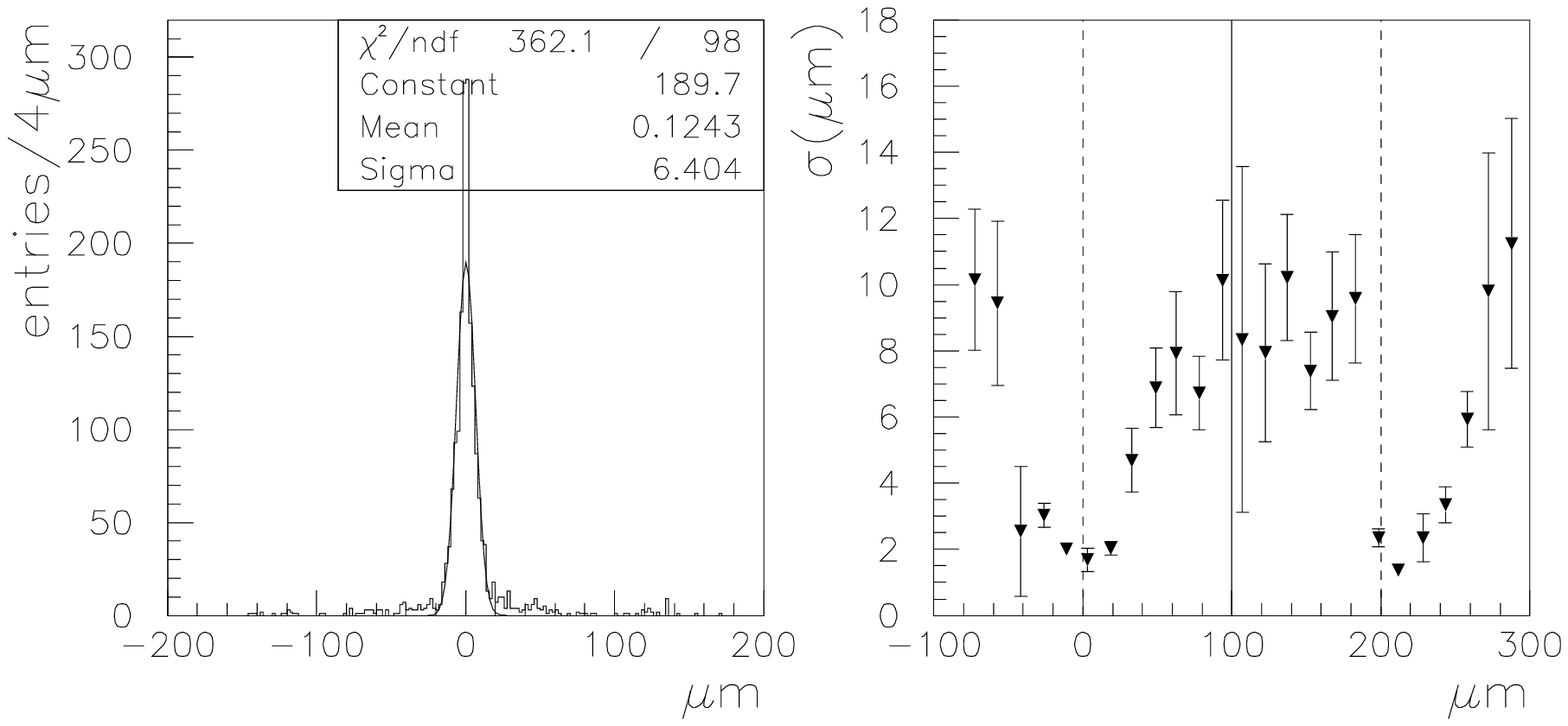,width=11.5cm,height=7.5cm,clip}
\caption{\sl The average resolution (left) and the achieved resolution vs. 
the spot position (right). The vertical lines identify the centre of the 
pixels; the full line indicates the reference pixel used for the $\eta$ 
calculation}
\end{center}
\label{fig:res}
\end{figure}

In the interleaved pixels, where the charge sharing is most efficient, the 
resolution approaches $\simeq 3~ \mu m$, irrespective of the $\simeq$ 40\% 
charge loss. On the other hand, when the charge sharing is minimal the 
resolution degrades to $\simeq 10~ \mu m$, even with a peak pulse height. 
According to these results, the binary resolution defined by the 
$implant~pitch/\sqrt{12}$ can be improved by about a factor 4 for a 
configuration where the ratio between the charge carrier cloud r.m.s. and the 
pixel pitch $\simeq$ 0.8. A similar scaling factor can be expected for a 
minimum ionising particle detected by a pixel sensor with $20-25 \mu m$ pitch,
as long as the peak $N/S \le 1/100$ and the charge loss is $\le 50\%$. This 
would lead to an intrinsic resolution $\simeq 2 \mu m$, matching well the
requirements discussed above.

\section{Discussion and Conclusions}

A Vertex Tracker based on a novel design of hybrid silicon pixel detectors 
is able to provide performances that are well adapted to the requirements
of the TESLA $e^+e^-$ collider while ensuring a significant safety margin 
against the influence of backgrounds. 
The ability to identify the individual bunch crossing of a $e^+e^-$ event 
reduces the occupancy and the confusion rate from pair background and, more
importantly, probability of overlap for a $\gamma \gamma \rightarrow 
{\mathrm{hadrons}}$ event. This ensures that the identification of processes
characterised by missing energy and/or heavy flavour jets at small polar angles
is not confused for this underlying background~\cite{gghad}.

The results reported in this paper conclude the feasibility study of the
proposed detector concept and indicate the direction for an intense 
dedicated R\&D programme towards a final detector design. This has to address 
several crucial issues:
\begin{itemize}
\item Sensor development. In order to achieve the desired resolution, 
a ${\rm 25\mu m}$ implant pitch detector is needed. The step from 
${\rm 50~to~25\mu m}$ is far from being a simple scaling. 
In fact, because of the limited inter-pixel 
space, biasing by suitable polysilicon resistors is not feasible. The most
promising solution is based on a punch-through mechanism, where very small 
${\rm p^+}$ dots along the metal lines of the bias grid serve sub-matrices of 
four pixels.
Moreover, the design has to be optimised with respect to the inter-pixel to
backplane capacitance ratio, with a target charge loss below 50\%.
A prototype run to validate the possible solutions is planned on a short 
time scale; this essentially concerns small test structures, meant to be 
read-out with a low noise microstrip detector electronics chip;

\item Electronics chip. On very general ground, each VLSI cell mating the 
pixel will integrate a low noise and low power charge sensitive amplifier, 
followed by a CR-RC shaper. 
Both blocks will be based on the folded cascade architecture 
with a PMOS transistor, chosen for its lower $1/f$ noise coefficient
at its input. The analog information of the pulse height could be efficiently 
translated into a time-over-threshold, locally digitised on every VLSI pixel 
cell. Because of the low duty cycle of TESLA, a pulsed mode 
operation seems very well possible, thus considerably reducing the power 
consumption. A readout system based on a matrix sparse data scan 
has to be included in the design, together with time stamping of each pixel 
hit. The shift-register architecture proposed for the 
ATLAS pixel detectors could be tailored to the linear collider requirements.
An end-of-column logical circuitry would deal with data packing and 
pipelining. The design of the chip should be optimised with the target of a 
cell dimension not exceeding ${\rm 75 \times 75 \mu m^2}$, to fully exploit 
the capacitive charge division mechanism. The development of a dedicated 
electronics chip being a considerable financial effort, this R\&D phase should 
be considered at an advance stage of the project approval process;

\item Technological development. Radiation hardening of the sensor and 
electronics design, back-thinning of the detector and chip, high
density interconnection (flip-chip), cooling, mechanics and system integration 
require an experimental effort to evolve from the present conceptual design to
a fully engineered study that should be planned at a later stage.
\end{itemize}

\section*{Acknowledgements}
The R\&D activity summarised in this report has been carried out in the 
framework of the $2^{nd}$ {\sl Joint ECFA/DESY Study on Physics and Detectors 
for a Linear Electron-Positron Collider} and it has been supported in part by
MURST under grant 3418/C.I. and by the Academy of Finland under the 
{\sl R\&D Program for Detectors at Future Colliders} grant.
We wish to thank R.~Brenner, P.~Jalocha, C.~Meroni, R.~Orava, H.~Palka, 
T.~Vanhala and K.\"Osterberg for their contributions at different stages of 
this activity and C.~Damerell for his suggestions.

\end{document}